# Monitoring the atmospheric throughput at Cerro Tololo Inter-American Observatory with aTmCam

Ting Li[*], D. L. DePoy, J. L. Marshall, D. Q. Nagasawa, D. W. Carona, S. Boada

Department of Physics and Astronomy, Texas A&M University, 4242 TAMU, College Station, TX 77843, USA

## ABSTRACT

We have built an Atmospheric Transmission Monitoring Camera (aTmCam), which consists of four telescopes and detectors each with a narrow-band filter that monitors the brightness of suitable standard stars. Each narrowband filter is selected to monitor a different wavelength region of the atmospheric transmission, including regions dominated by the precipitable water vapor and aerosol optical depth. The colors of the stars are measured by this multi narrow-band imager system simultaneously. The measured colors, a model of the observed star, and the measured throughput of the system can be used to derive the atmospheric transmission of a site on sub-minute time scales. We deployed such a system to the Cerro Tololo Inter-American Observatory (CTIO) and executed two one-month-long observing campaigns in Oct-Nov 2012 and Sept-Oct 2013. We have determined the time and angular scales of variations in the atmospheric transmission above CTIO during these observing runs. We also compared our results with those from a GPS Water Vapor Monitoring System and find general agreement. The information for the atmospheric transmission can be used to improve photometric precision of large imaging surveys such as the Dark Energy Survey and the Large Synoptic Survey Telescope.

## 1. INTRODUCTION

The discovery of the accelerating Universe is one of the most important discoveries in Cosmology. Within the framework of the standard cosmological model, this acceleration implies that about 70% of the Universe is composed of a dark energy component, for which there is no persuasive theoretical explanation. A variety of projects are underway or planned to more fully explore dark energy parameters. For example, the Dark Energy Survey (DES) project plans to use a variety of techniques to estimate the dark energy equation of state parameter, $w$.[1] DES is underway using the Dark Energy Camera (DECam) at the Cerro Tololo Inter-American Observatory (CTIO) 4m Blanco telescope.[2,3] The Large Synoptic Survey Telescope (LSST) project also has dark energy related science objectives.[4] These projects use multi-color imaging surveys of large areas of the sky at optical wavelengths to produce catalogues and light curves of objects, which are in turn used to probe the parameters of the Universe. Crucially, the photometric precision of the surveys must be excellent, so as to avoid significant systematic errors in the results. The systematic uncertainty of the measurements of $w$ from the first three years of the Supernova Legacy Survey (SNLS3), for example, is dominated by the photometric precision of the survey.[5] Of course, many other science projects demand exquisite photometric precision; see the "LSST Science Book" for many examples.[4]

DES has an overall photometric precision goal of ~0.01 mag in the primary survey bands; LSST's goals are tighter at ~0.005 mag. Although traditional photometric observing techniques (numerous standard star measurements, observations over a large range of airmass, selection of standard stars with a range of colors, etc.) can produce photometric precision at these levels, assembling a large survey with such precision is very challenging. For example, the Sloan Digital Sky Survey (SDSS) achieved relative photometric precision of roughly 1-2.[6] The source of precision error in large surveys is thought to be a combination of various calibration uncertainties. One of these is related to the exact bandpass and response function of the imaging system (including optics, filters, detectors, etc.); DES has deployed a system that calibrates the instrumental throughput with excellent precision.[7] Figure 1 shows the system response function of DES ugrizY bandpasses measured by this calibration system. But it is believed that the largest source of photometric precision error is due to variations in atmospheric throughput. For example, Ivezić et al. (2007) found that assuming a standard atmosphere can induce 0.01 mag offsets in some colors due to differences in the real atmosphere during observing relative to the assumed standard.[8] Stubbs et al. (2007) also show that expected changes in the transmission of the atmosphere can produce 1- 2% photometric precision errors.[9]

*sazabi@neo.tamu.edu

Atmospheric transmission in the optical wavelengths (~300nm-1100nm) is mainly determined by 3 processes: Rayleigh scattering from molecules, aerosol and dust scattering from small particles, and molecular absorption (principally by $O_2$, $O_3$, and $H_2O$). Figure 1 shows a model of fiducial atmospheric transmission at CTIO at airmass X=1.3. Another water-free atmospheric model is also plotted as contrast to see particularly the absorption from $H_2O$.

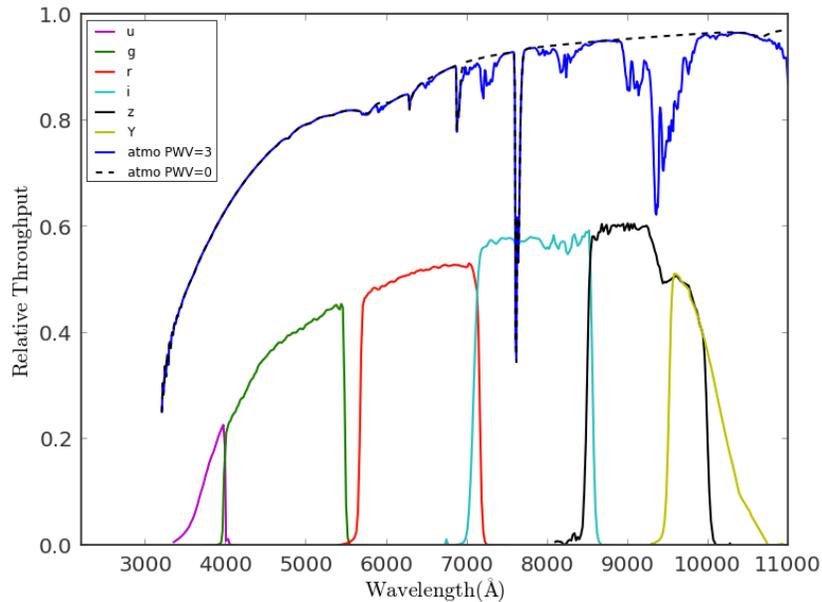

Figure 1: DES filter bandpasses (ugrizY) and model of atmospheric transimission. A model of fiducial atmospheric transmission at CTIO is shown, with altitude 2.24km, barometric pressure P = 780mbar, aerosol optical depth tau=0.05 at 550nm, and precipitable water vapor PWV=3mm at airmass X=1.3, generated by libRadTran. Another PWV-free atmospheric model is plotted as contrast to see where the absorptions from $H_2O$. For reference, the system response function of ugrizY bandpasses currently installed in DECam are shown.

The throughput of the atmosphere has been measured by astronomers in great detail using spectroscopic observations. For example, Wallace et al. (2007) gives a high resolution transmission spectrum of the atmosphere from Kitt Peak National Observatory obtained during the course of observing the Sun.[10] Of course, characterization of the atmosphere is of general interest and is the specific interest of a wide range of scientists. There are, for example, many atmospheric modeling packages available: e.g., MODTRAN[11]; libRadtran[12]; ATRAN[13]. These packages are used to aid interpretation aimed at projects such as atmospheric ozone concentration determination, pollution monitoring, solar irradiance measurements, meteorological effects on agriculture, and other areas of interest to the atmospheric science community. Sometimes, astronomical measurements are of direct interest to these communities.[14,15]

In general, astronomical spectroscopic determination of the atmospheric transmission proceeds by observing a specific target (or a set of targets with well-understood SEDs) at various airmasses. In particular, spectroscopic observations of standard stars at ~5 minute cadence, over a range of airmasses, and at wavelengths of $400 < \lambda < 1000$nm can produce high quality atmospheric absorption profiles. While this approach is ideal, it has a major drawback in requiring a high level of personnel commitment to aligning a relatively small aperture (~10 arcsec) on the target stars, a relatively large telescope, and a stable spectrograph.[16]

Although this technique produces reasonable results, there have been relatively few systematic, long-term studies of the detailed atmospheric transmission at any major astronomical observatory. Indeed, very little is known about the time- and angular-scales of changes in atmospheric transmission from an astronomical perspective. This is particularly true for

"non-photometric" nights, where (quite reasonably) there are generally no attempts to report (or measure) extinction coefficients.

In a previous paper[17] (Li et al. 2012, hereafter L12), we described a simple system that rapidly monitors the transmission of the atmosphere. The Atmospheric Transmission Monitoring Camera, or aTmCam, uses simultaneous measurements of stars with known spectral energy distributions through a set of narrow-band filters. The filters are chosen to allow determination of specific features in the atmospheric transmission spectrum, which then can be used to develop a model that accurately represents the throughput of the atmosphere. The system is similar to that suggested by Stubbs et al. (2007). L12 used a concept testing system to demonstrate that brightness measurements of stars at a few wavelengths can be used to derive a model for the transmission of the atmosphere that is as precise as what can be derived with spectroscopic measurements. L12 coupled a small telescope and a commercial spectrometer to obtain spectroscopic measurements of bright stars, simultaneously measured the brightness of the same star through a set of narrow-band filters, and showed that the atmospheric transmission derived from both measurements is the same. In this paper, we present the work that has been done with a prototype version of aTmCam based on the principles similar to the imaging system proposed in L12. We have used the prototype at CTIO for ~40 nights of observing in 2012 and 2013 and have determined (over these particular nights) the angular and temporal scale of meaningful changes in the atmosphere. In Section 2, we describe our system and instruments. This is followed by a description of the observations and data reduction in Section 3. The results are presented in Section 4, followed by a conclusion in Section 5.

## 2. DESCRIPTION OF THE SYSTEM

We have built and deployed an atmospheric transmission monitoring system (aTmCam) based on simultaneous measurement of the brightness of a star (with well-understood SED) through narrow-band filters using a well-calibrated system. L12 give additional details of the system design, configuration, operation and data reduction. The aTmCam currently consists of four Celestron $f/10$ 8-inch telescopes mounted on two Celestron CGEM mounts. Each telescope is fitted with an SBIG ST-8300M CCD and a filter centered near a part of the spectrum sensitive to a particular component of the atmospheric throughput, with a different filter in each telescope. The pixel scale of the system is about 0.54"/pixel and the Field of View (FOV) is about 20' x 30'. Figure 2a shows a photograph of (half of) the prototype being deployed at CTIO in October 2012. The central wavelengths of the four filters we used here are 390nm, 520nm, 852nm and 940nm as shown in Figure 2b. The details of the bandpasses, part numbers, etc. can be found in L12. There were initially 5 filters in our original study, as noted in L12; we eliminated the filter with central wavelength of 610nm in this work because the ozone absorption around 610nm (known as the Chappuis band) is small and not sensitive to the variation of the ozone column density. The photometric error produced by ozone variation in Chappuis band is also negligible. The ozone absorption mostly takes effect below 320nm and DES has no specific interest at that wavelength range. As shown in Figure 1, the DES u band has almost no throughput below 320nm. However, ozone variation will be important for the LSST since the LSST u band filter has significantly higher throughput and extends to a bluer wavelength.

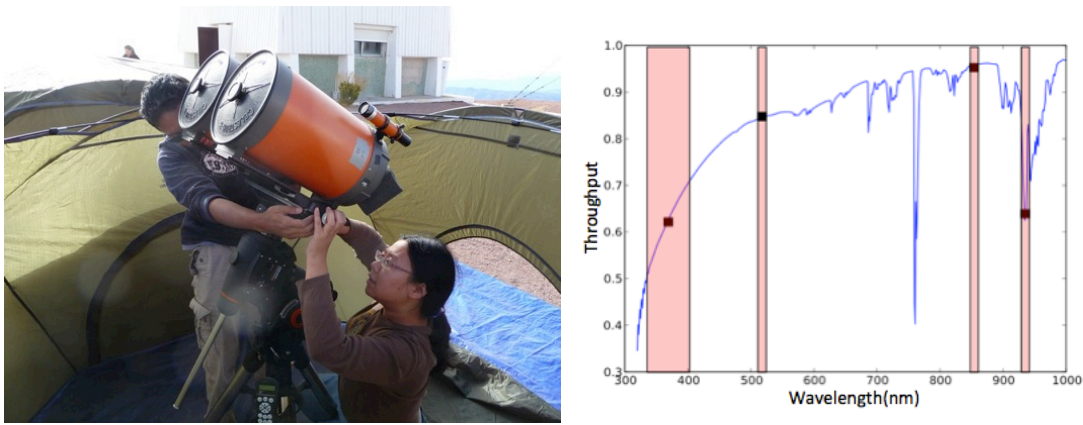

Figure 2a. Photograph of one of the prototype systems being installed at CTIO in November 2012. The prototype used a tent with a removable top as a temporary shelter during the observing campaign. Figure 2b. The central wavelengths of the four filters we used in this study overplotted on a fiducial atmospheric model.

Apart from the hardware, we have also developed our own control software under the OS X operating system using Apple's native object-oriented application programming interface *Cocoa*. The software can command the two mounts and four CCDs so that they can point at one star and take simultaneous images in four filters. Simultaneous observations of the brightness of a star with each telescope allow determination of "colors" that measure the atmospheric transmission, as described in the next section.

## 3. OBSERVATION AND DATA REDUCTION

We had two one-month-long observing campaigns with our prototype instrument as described above at Cerro Tololo Inter-American Observatory (CTIO). The first observing run was conducted during October-November 2012 during the Dark Energy Survey (DES) Scientific Verification (SV) season. The second run was operated during September-October 2013 during the DES Year 1 Observation season. We spent a few days and nights to set up our system and instrument, including polar alignment with drifting method to ensure tracking accuracy within 1"/min, as well as multi-star pointing to ensure pointing accuracy within 5' (about quarter of the FOV). We lost only a few nights due to the bad weather (cloudy nights or storms) during these observing campaigns. We conducted 14 nights and 21 nights of successful observation in 2012 and 2013, respectively. The main objective of these two campaigns was to characterize typical timescales and angular scales of significant changes in the atmospheric transmission above a high quality astronomical site. We therefore ran the system in two different modes. We spent several nights continuously observing a single star (generally HIP117452) as it moved across the sky to study the temporal variation of the atmosphere (hereafter temporal mode). The star was selected to enable a total of ~10 hours of observation every night and ~5 hours before and after crossing the meridian. We also spent many nights pointing the system at stars around the sky at numerous positions to study the angular variation of the atmosphere (hereafter angular mode); the system is capable of slewing between targets in ~5 minutes. All the targets we selected are bright B7V-A1V stars with $3 < V < 6.5$ mag from the Hipparcos Catalog. In both modes, we took simultaneous images in four filters. Since the CCD QE at 520nm is about 10 times better than that at 940nm, and also the target stars also have much higher flux at 520nm, the exposure time could not be the same for all 4 CCDs. In order to ensure the 520nm band was not saturated and the 940nm band had adequate signal-to-noise ratio (S/N), we set the exposure time to be 1, 1, 5, 5 seconds for 380nm, 520nm, 852nm and 940nm, respectively. For the nights the sky was covered by a thin layer of clouds, we adjusted the exposure time as needed. Also, the exposure time was varied from target-to-target to avoid saturation and obtain adequate S/N. The typical S/N was larger than 50 for band 380nm, 520nm, 852nm and about 20-30 for band 940nm. It takes about 20 seconds for 4 CCDs to read out, so we took about 2 exposures per minute in temporal mode and we obtained about 1000 images per filter per night. In angular mode, we took about 10 exposures on each star and move to the next one with about a 5-minute overhead. We obtained roughly 500 images per filter per night in angular mode. Every night, we adjusted the telescope focus 2-3 times as the temperature changed.

All the data were bias subtracted with the bias frames taken every day before the start of observation. Then the images were flattened using frames taken during the evening twilight when the sky is clear. If the sky flats were not taken due to the partly cloudy weather, then the sky flats from the previous night were used.

The instrumental magnitude for each narrow-band image was measured using aperture photometry with IRAF[a]. Since the telescope focus is sensitive to the temperature and the image PSFs changed significantly over one night, we used a large aperture (~7") to ensure all the stellar flux was measured equally in all frames, which minimizes systematic photometric errors due to seeing or PSF variation. We refer to the instrumental magnitudes from the four narrow bands as $m_{380}, m_{520}, m_{852}$ and $m_{940}$.

We generated a grid of models of the Earth's atmospheric transmission with different column densities of precipitable water vapor (hereafter PWV), aerosol optical depth (AOD) at 550nm, $\tau_{550}$, and barometric pressure using libRadTran. We calculated the synthetic color with those models, the expected throughput of the system and the SED of an A0V star.

---

[a]IRAF is distributed by the National Optical Astronomy Observatory, which is operated by the Association of Universities for Research in Astronomy, Inc., under cooperative agreement with the National Science Foundation

Then a "best fit" model was determined by minimizing $\chi^2$ between the model and data values. A unique PWV and AOD is taken from the best fit model.

For example, the $m_{852} - m_{940}$ color index is mostly sensitive to PWV but not other atmospheric parameters. We therefore could derive the PWV from the $m_{852} - m_{940}$ color. The left panel of Figure 3 plots the synthetic color index $m_{852} - m_{940}$ and airmass X at various PWV but a given pressure and AOD. The color is proportional to $X^{0.58}$ and converges at X=0. Using this relationship, we could derive the PWV by measuring the color at a given airmass. Especially, on the nights that the PWV was stable, the slope also allows estimation of the PWV of that night as a consistency check.

AOD may be determined using the $m_{520} - m_{852}$ color index. A relation between the synthetic color index $m_{520} - m_{852}$ and airmass X at various AOD and a given PWV and pressure is shown in the right panel of Figure 3. Note that $m_{520} - m_{852}$ is proportional to X while $m_{852} - m_{940}$ is proportional $X^{0.58}$. This is because $m_{940}$ is centered on a wavelength at which the strength of the $H_2O$ absorption feature is not linear with the $H_2O$ column density.

In principle, we could derive the pressure in a similar way as PWV and AOD using $m_{380} - m_{520}$, but we instead obtain the barometric pressure from the weather monitoring station at the site on CTIO.

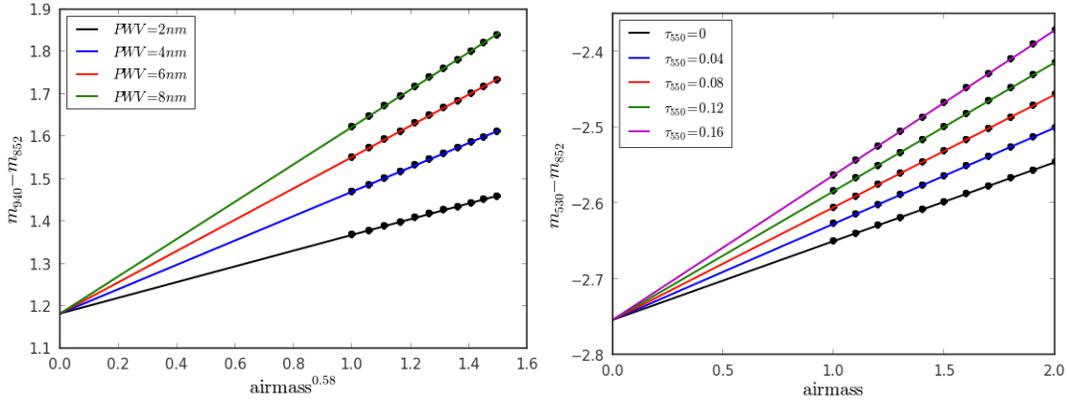

Figure 3. Synthetic color indices vs. airmass at various PWV (left) and AOD (right).

## 4. RESULTS AND DISCUSSION

### 4.1 Clouds are grey

As mentioned previously, the best atmospheric models are determined by the colors of the stars measured with a set of narrow-band filters and there is a one-to-one relation between the measured colors and each atmospheric component such as PWV, AOD, etc. Some of our measurements were conducted under non-photometric conditions, i.e. the sky was partly covered by the clouds. The assumption generally is that clouds are grey and so will not affect measured photometric color indices. However, the aTmCam system can be used to demonstrate unambiguously that clouds are indeed grey.

On the night of Sep 25, 2013, the sky was partly cloudy for the first half of the night and later cleared. We spent the entire night monitoring one star HIP 117452 as it crossed the sky. We can therefore derive an airmass extinction term for each filter using the data from the second half of the night as show in the top left panel of Figure 4. The bottom left panel of Figure 4 shows the airmass extinction-corrected magnitude in filter 520nm, $m'_{520}$, as a function of time. At the beginning of the night, the clouds caused as much as 1.5 mag of extinction. We also calculated the color index after the extinction correction, $m'_{380} - m'_{520}$, as a function of time, shown in the bottom right panel. $m'_{380} - m'_{520}$ is constant on average over time, suggesting that clouds are grey. The cyan line shows the 1-σ scatter from the mean. Though the color variation at the beginning of the night is as large as 0.1 mag, it is still consistent with expected S/N calculation as the noise was dominated by the background noise (mainly the read-out noise) when the clouds caused large extinction. We did a similar test on the other bands and all show similar results. We thus conclude that clouds are grey.

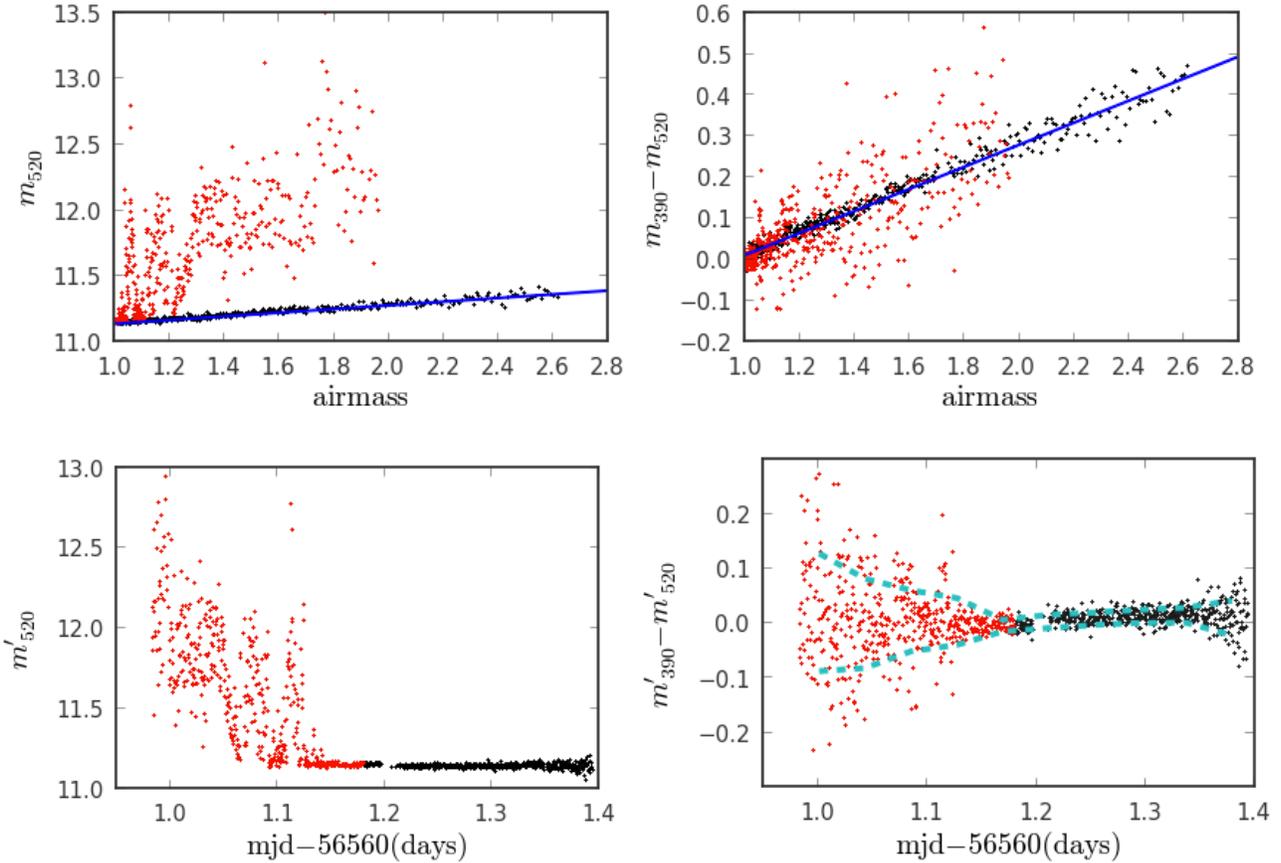

Figure 4. Top left (right) panel: Instrumental magnitude $m_{520}$ (color index $m_{380} - m_{520}$) vs. airmass measured on night of Sep 25 2013. Red dots show the measurements from the first half the night and black dots are from the second half of the night. A linear fit is applied only on the measurement from the second half of the night to derive an airmass extinction term for that band. Bottom left (right) panel: The airmass extinction corrected magnitude $m'_{520}$ (color index $m'_{380} - m'_{520}$) vs. time. The cyan line shows the 1-σ scatter from the mean.

### 4.2 Results of PWV and AOD

For each measurement obtained during both observing runs, we derive a best-fit AOD and PWV. We show five nights of the results from the 2013 observing run in Figure 5. It can be seen that the PWV (left column) can change by a few millimeters over one night and mostly decreases over time on any given night. The AOD (right column) is stable and below 0.05 for entire run except the night of Sept. 21 and Sept. 23, which is shown in the Figure 5. The scatter is mainly the statistical error due to the photon noise. The 1-σ error is ~0.6mm in PWV and ~0.03 in AOD. Overplotted in Figure 5 is the PWV measured by a GPS Water Vapor Monitoring System (hereafter GPS), a fully automated dual-band GPS ground station mounted outside the CTIO-1.5m telescope dome[b]. The PWV measured with two independent methods agrees within the joint uncertainties of the two measurements. The difference is generally less than 1 mm, which agrees with the errors of the GPS measurement (shown as errorbars in Figure 5). However, the PWV measured by the GPS can have large errors at the beginning of each night. Also, GPS inexplicably failed to measure the PWV for some nights so we cannot compare the results for the night of Sept 20 and 21, 2013, for example.

---

[b] http://www.suominet.ucar.edu/index.html

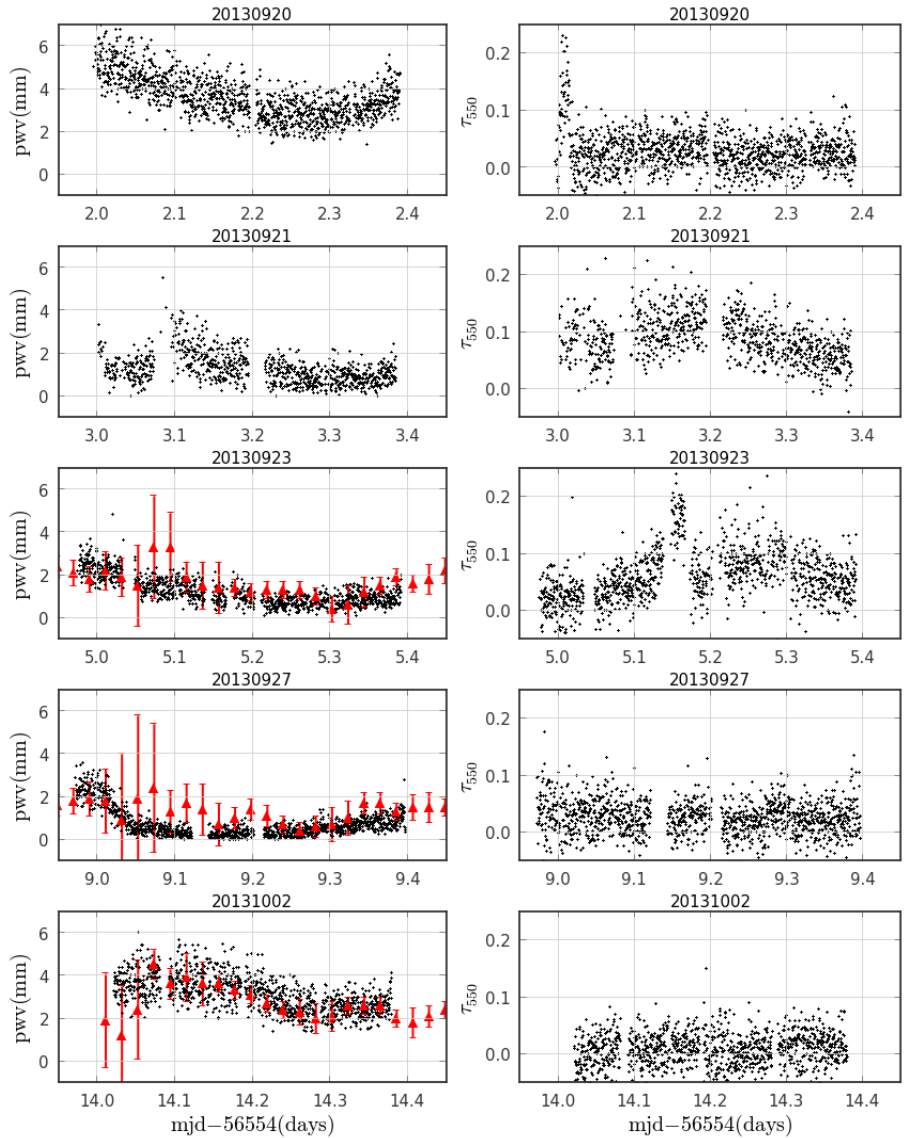

Figure 5: An example of 5 nights of the results from 2013 observing run. The left (right) column shows the PWV (AOD) as a function of time in MJD. The number on the top of each panel is the date of the night that the observation started, in the format of YYYYMMDD. Overplotted red triangles are the PWV measured by a GPS Water Vapor Monitoring System.

Figure 6 plots the normalized probability density function (PDF) of PWV measured during both observing runs. The PWV varies between 0 and 6mm with a median of ~2mm for both runs. However, the GPS measured the PWV to be >10mm a few nights after our observation run in 2013. We believe that more observations with aTmCam at other seasons throughout the year will help us to understand better the full range of potential values of PWV at CTIO.

Since the PWV is unstable compare to the AOD, we studied the variation of PWV in detail in the following sections.

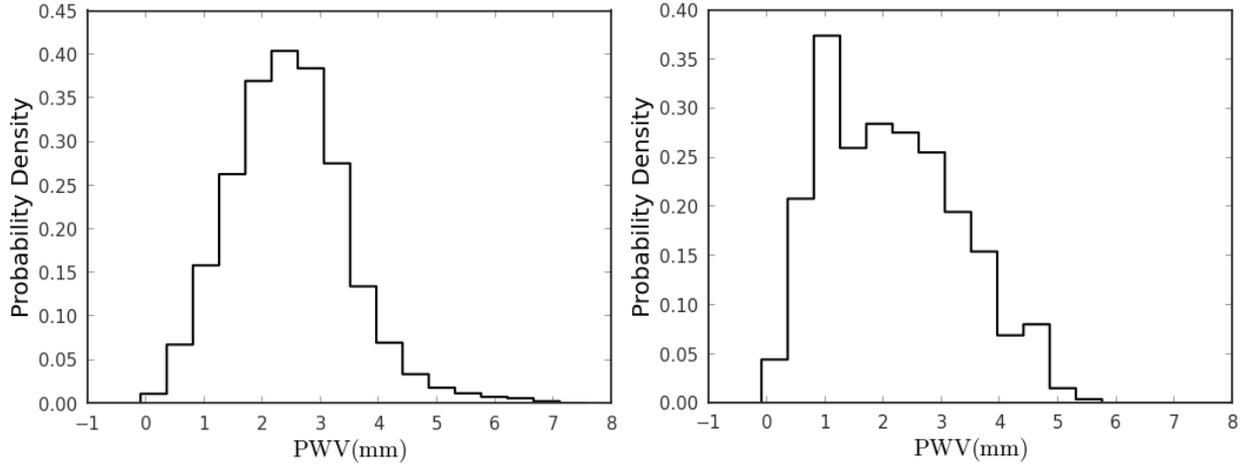

Figure 6: Normalized probability density function (PDF) of PWV measured in 2012 observing run (left) and 2013 observing run (right).

### 4.3 Temporal Variation of PWV

A very important purpose of this study is to understand how rapidly the characteristics of the atmosphere that affect astronomical photometric measurements change. Here we studied the time scale of the PWV variation with the thirteen nights of observations from the 2013 campaign, during which we observed a single star as it moved across the sky. We averaged the PWV measurements with a binning size of 3 minutes. Then we calculated the change of the PWV, $\Delta PWV$, as a function of time scale, $\Delta t$, as follows:

$$\Delta PWV(\Delta t) = PWV(t + \Delta t) - PWV(t)$$

The normalized probability distribution function (PDF) of $\Delta PWV$ in 3 minutes, 1 hour, 3 hours, and 6 hours is shown in Figure 7. As the time scale gets larger, the PDF shifts towards left and the width gets broader. The mean and the standard deviation of $\Delta PWV$ at different time scales are plotted in Figure 8. Similar to what is seen in Figure 7, PWV decreases by about 1mm over ~9 hours on average. The standard deviation of measured $\Delta PWV$, $\sigma(\Delta PWV)'$, is a combination of the actual standard deviation of $\Delta PWV$ and measurement errors due to the photon noise, i.e.:

$$\sigma(\Delta PWV)' = \sqrt{\sigma(\Delta PWV)^2 + \sigma(\text{measurement})^2}$$

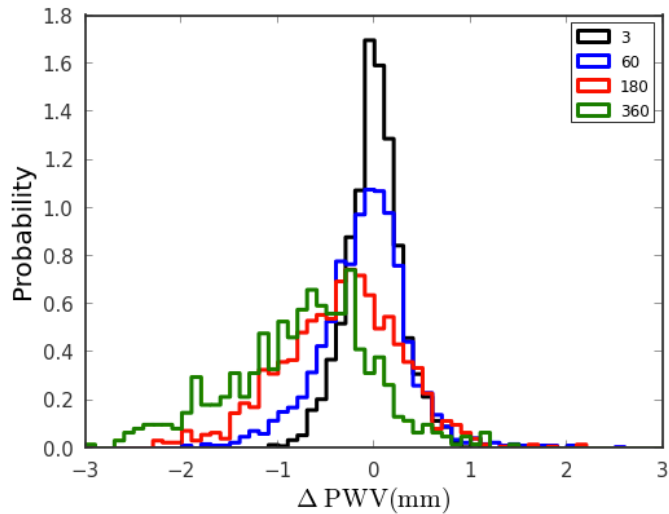

Figure 7: The distribution of $\Delta PWV$ in 3 minutes, 1 hour, 3 hours, and 6 hours.

If we assume that the atmosphere is stable over 3 minutes and σ(ΔPWV)′ at 3 minutes is only due to the measurement errors,

$$\sigma(\text{measurement}) = \sigma(\Delta \text{PWV}|_{3\text{minutes}}) = 0.29 \ mm$$

After subtracting the measurement errors, we could calculate the broadening just due to the variation of the PWV, which is shown as the red dots in Figure 8. That is, the probability of the change of PWV, or P(ΔPWV), after time Δt, will have a distribution with mean μ(ΔPWV) and standard deviation σ(ΔPWV), which is shown in Figure 8.

Now we can use the character of this temporal variation to study how fast we need to measure the PWV. Assuming that P(ΔPWV) has a Gaussian-like distribution, we could calculate the probability of the PWV changing by an amount less then $P_0$, or $P(|\Delta PWV| < P_0)$, as a function of time. In Figure 9, we show the case for $P_0 = 1mm$ and $P_0 = 0.5 \ mm$. In an hour, there is a 99% (80%) chance that the change of PWV is less than 1mm (0.5mm). If we are only concerned with measuring the PWV with an accuracy of 1mm, then measuring the PWV once per hour is sufficient.

We obtained similar results during the 2012 campaign, which suggests that these timescales are generally pertinent for CTIO (at least for these months).

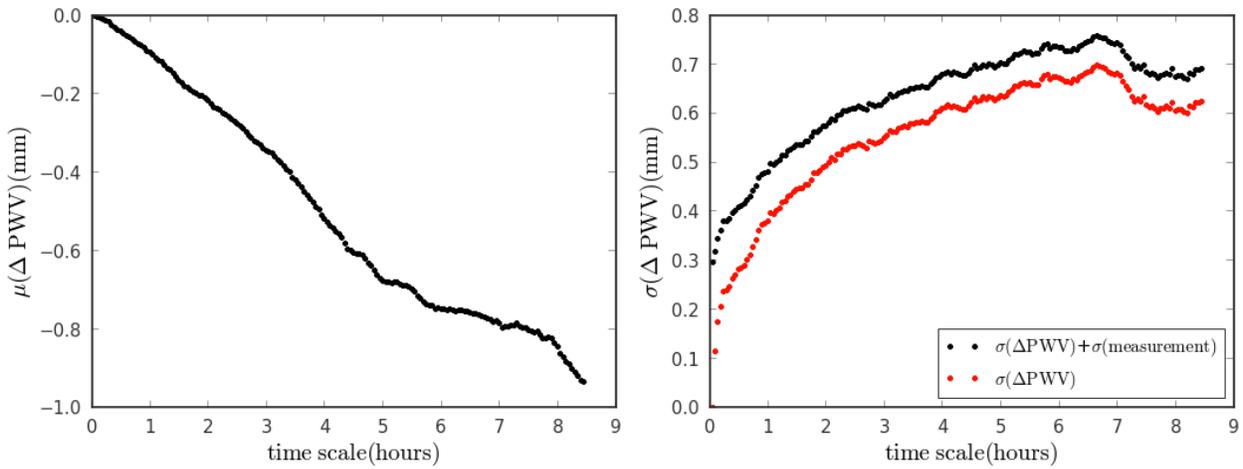

Figure 8: The mean (left) and the standard deviation (right) of the PDF of ΔPWV at different time scales. The red dots in the right panel show the actual standard deviation of ΔPWV after the measurement errors are removed.

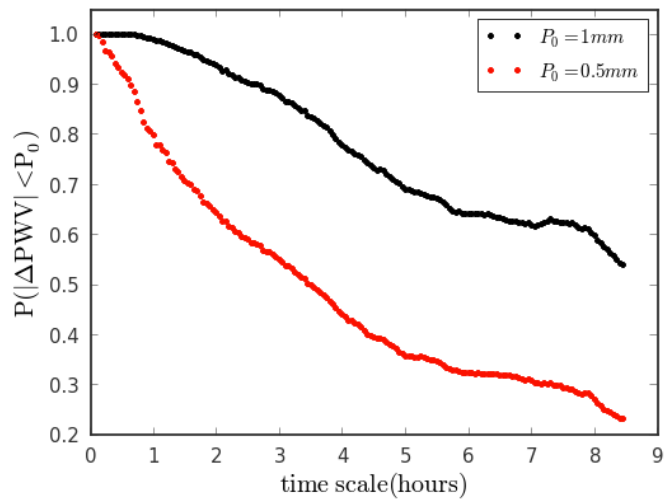

Figure 9: $P(|\Delta PWV| < P_0)$ as a function of time. In an hour, there is a 99% (80%) of the chance that the change of PWV is less than 1mm (0.5mm).

## 4.4 Angular Variation of PWV

Another key question of this study is to understand how uniform the atmosphere is across the sky. We therefore spent 8 nights to study the angular variation of the PWV by measuring a grid of A0V stars all over the sky during the 2013 campaign. In Figure 10, we show the telescope pointing in Hour Angle (HA) and Declination (Dec) for one night of observing. For each pointing, we took about 10 exposures in 5-6 minutes and then slewed to the next star with ~3-5 minutes overhead time. For each star, we averaged the PWV into 2 measurements with a binning size of ~3 minutes.

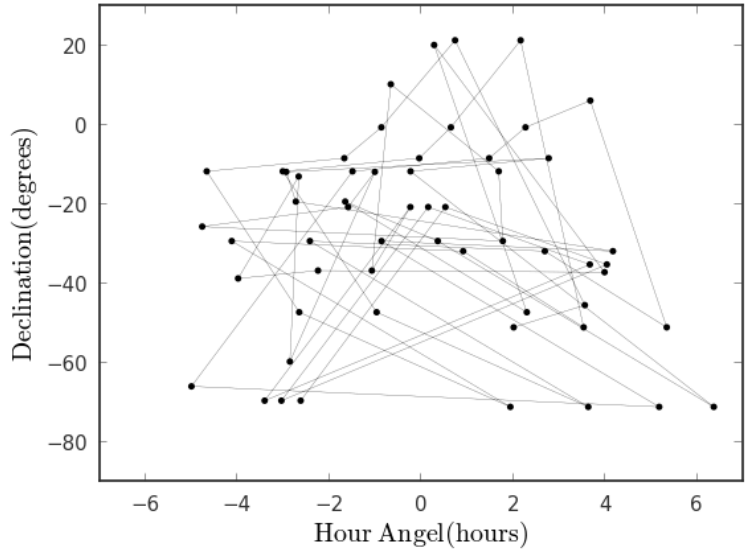

Figure 10: Telescope pointing in HA and Dec for a grid of A0V stars taken during the night of Oct 09 in 2013 observing run. The points represent the HA/Dec of the stars when the images were taken, and the solid lines trace the temporal order of the observation.

We then calculate the change in PWV, or ΔPWV, in two consecutive measurements,

$$\Delta PWV = PWV_{k+1} - PWV_k$$

We assume that the atmosphere has no temporal variation between the two measurements and the change is solely due to the angular variation. We plot ΔPWV as a function of the separation angle θ of two measurements in Figure 11. There are a group of points near θ=0 that corresponds to the repeated measurements of the same star before the telescope slewed to the next object. The distribution of these points also gives an estimate of the measurement error of 0.3mm, similar to that measured in the temporal variation study. The remaining points are computed from the last measurement taken at one pointing and the first measurement taken after a slew of the telescope. We also average the ΔPWV with a binning size of 20 degree in separation angle (except that the first two bins are 0-3 degrees and 3-20 degrees.), shown as red open circles. The error bars in Figure 11 show the 1-σ errors for each bin. It can be seen that this scatter is stable (and mainly due to measurement error) up to ~90 degrees across the sky, which indicates that the PWV was angularly homogeneous during the 2013 observing run.

Shown in Figure 12 is the change in PWV plotted against the change in HA and Dec. Positive in HA (Dec) represents the telescope slew from East (South) to West (North). We saw no trend in an east-west or north-south gradient in PWV during the 2013 observing run.

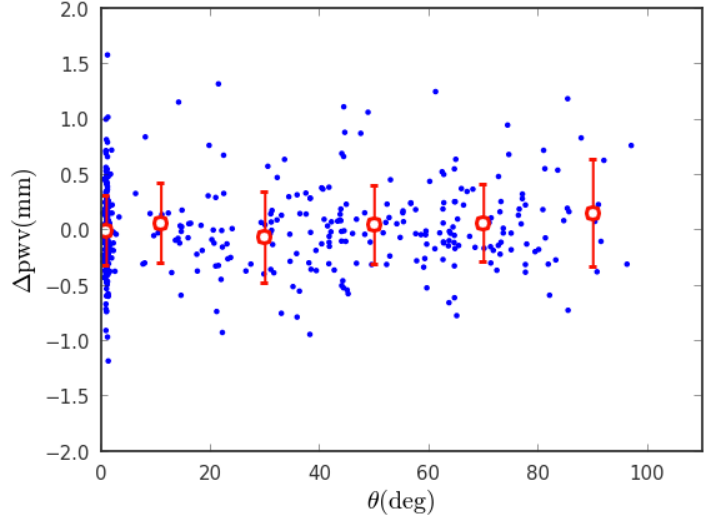

Figure 11: Angular variation of PWV at the CTIO site for the 2013 observing run. Blue dots represent the change in PWV; red open circles are the average PWV with a binning size of 20 degree (except that the first two bins are 0-3 degrees and 3-20 degrees). The error bars are the 1-σ scatter in each bin. This scattering is stable (and mainly due to measurement error) up to ~90 degrees across the sky, which suggests that the sky was homogeneous during our run.

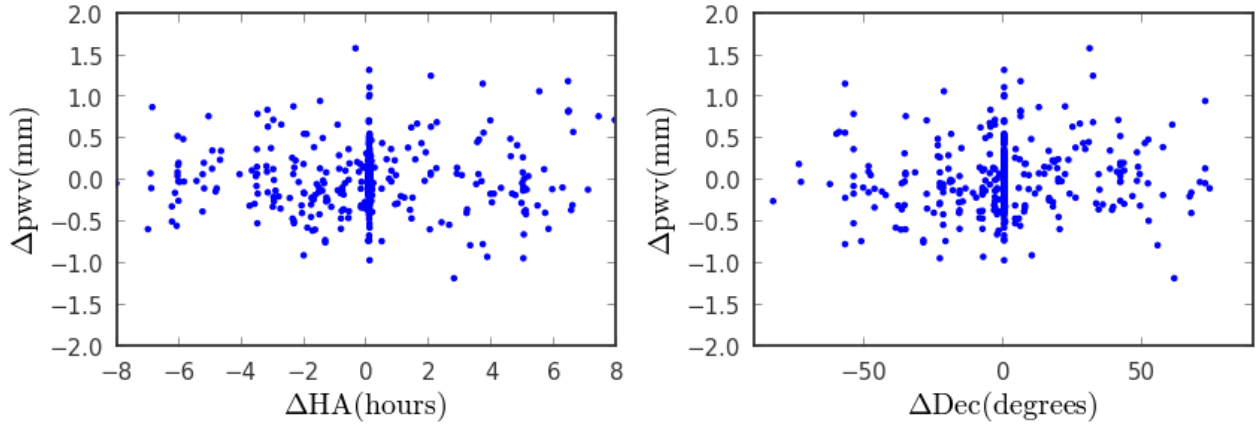

Figure 12: Angular variation of the PWV as a function of east-west (left) and north-south (right) pointing of the telescope. No obvious trend of an east-west or north-south gradient was found during the 2013 observing run.

## 5. CONCLUSIONS

We present results from a simple system that has been used to monitor the transmission of the atmosphere above the CTIO site. This system, aTmCam, consists of four telescopes and detectors each with a narrow-band filter that monitors the brightness of suitable standard stars simultaneously. We deployed this system to CTIO for ~40 nights of observing in 2012 and 2013 and we have derived the precipitable water vapor and aerosol optical depth from the measured color of the stars. We achieve a precision of ~0.6 mm of PWV and ~0.03 in AOD. We see that the precipitable water vapor can change over one night (typically decreasing), while aerosol optical depths are generally quite stable. We probe the time and angular scales of the variation of the precipitable water vapor. During our observing runs, we conclude that we need to measure the PWV only once per hour if we require PWV estimates accurate to 1mm. We also observe no significant PWV variation over an angle of ~90 degrees on the sky. Again, these measurements are consistent over two observing seasons, albeit during the same general time of the year.

We show that aTmCam is a simple and capable system to monitor the atmospheric transmission. It will be of general interest to have more such observations at CTIO (or Cerro Pachon for LSST) to fully sample a wide range of atmospheric conditions and other seasons. The information for the atmospheric transmission, if obtained simultaneously with survey operations, could be used to improve the photometric precision of large imaging surveys such as the Dark Energy Survey and the Large Synoptic Survey Telescope.

## ACKNOWLEDGEMENTS

Texas A&M University thanks Charles R. '62 and Judith G. Munnerlyn, George P. '40 and Cynthia Woods Mitchell, and their families for support of astronomical instrumentation activities in the Department of Physics and Astronomy. The authors would also like to thank the support staff at CTIO for their assistance.